\documentclass[10pt,journal,compsoc]{IEEEtran}
\usepackage[T1]{fontenc}

\ifCLASSOPTIONcompsoc
  \usepackage[nocompress]{cite}
\else
  \usepackage{cite}
\fi
  \usepackage[dvips]{graphicx}
   \graphicspath{{../eps/}}
  \DeclareGraphicsExtensions{.eps}
\usepackage{algorithmic}

 \usepackage[caption=false,font=footnotesize,labelfont=sf,textfont=sf]{subfig}
\usepackage[caption=false,font=footnotesize]{subfig}
\begin{document}
%
\title{A hybrid chaos map with two control parameters to secure image encryption algorithms}
%
%
%
%

\author{Roghayeh~Hosseinzadeh,~
        Yavar~Khedmati,~
        and~Reza~Parvaz
\thanks{R. Hosseinzadeh, Y. Khedmati and R. Parvaz were with the Department
of Mathematics, University of Mohaghegh Ardabili,Ardabil 56199-11367, Iran
.}
}
\markboth{Journal of \LaTeX\ Class Files,~Vol.~14, No.~8, August~2015}%
{Shell \MakeLowercase{\textit{et al.}}: Bare Demo of IEEEtran.cls for Computer Society Journals}
%



\IEEEtitleabstractindextext{%
\begin{abstract}
In this paper, we introduce a hybrid chaos map for image encryption method with high sensitivity.
This new map is sensitive to small changes in the starting point and also in control parameters which result in having more computational complexity. Also, it has uniform distribution that provides resisting of the new system against attacks in security applications. Various tests and plots are demonstrated to show more chaotic behavior of the proposed system. Finally, to show the ability of the generated chaotic map in the existences image cryptography approaches, we further report some results in this area.

\end{abstract}

}

\maketitle


\IEEEdisplaynontitleabstractindextext

%
\IEEEpeerreviewmaketitle

\IEEEraisesectionheading{\section{Introduction}\label{sec:introduction}}


%
%
%
%
\IEEEPARstart{D}{ue} to increasing demand for information security, cryptography is used to transfer information safely.
 Chaos is known as a natural tool for cryptography applications because of its properties such as unpredictability and initial state sensitivity.

  Early chaotic systems such as logistic
and tent maps have weaknesses such as limited chaotic area and non uniform
distribution. The idea of combining these systems and creating
new systems has been developed to solve the mentioned problems \cite{1,2,3}.  A combined chaotic system has been suggested and applied to color image encryption in \cite{4}. Also, a novel type of uniformly distributed 2D hybrid chaos map useing the cellular automata and discrete framelet transform has been developed to transfer images securely by using cryptography and steganography methods \cite{5}.

 The purpose of this paper is to introduce a new type of uniformly distributed hybrid chaos
  map useing chaos maps introduced in \cite[5,6]. The proposed system depends
on the values $r_1$, $r_2$ and the starting point unlike the existing systems which based on only the value $r$ and the starting point.
This idea leads to have more computational complexity, a large enough
positive Lyapunov exponent which provides hyper-chaos map, better sensitivity and high security.

Performance analysis are provided to show that this map has the wider chaotic range and hyperchaotic features than existing chaotic systems. To show its performance in security applications, it is utilized in the image encryption method. The image encryption algorithm introduced in \cite{7} is used to test the featurs of the new system.



\section{Hybrid chaos map}
In this section, the new chaos map is introduced using hybrid chaos and 2D hybrid chaos maps \cite{5,6}. Fristly, we recall the structure of these maps and then the new map is detailed in \ref{sec1}.
\begin{figure*}[!t]
\centering
\subfloat{\includegraphics[width=6.5in]{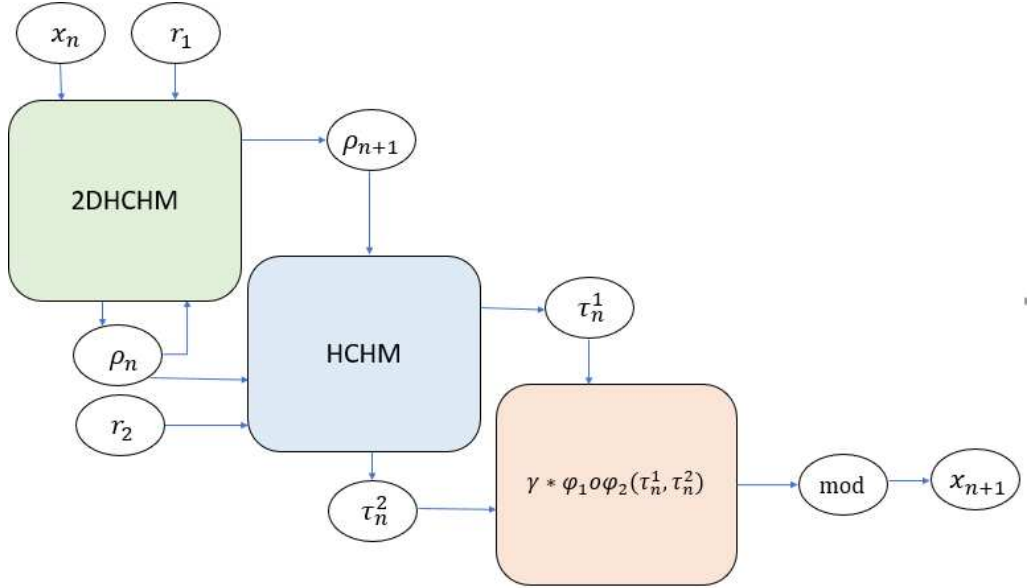}%
}
\vspace*{-1cm}
\caption{Proposed chaos map.}

\label{flo}
\end{figure*}
\subsection{Related work}
Hybrid chaos and 2D hybrid chaos maps are combination of Sine and Logistic Tent maps.
The general structure of Sine map is given as
$$x_{n+1}=s(r,x_n):=r \sin(\pi x_n)/4,$$
and Logistic Tent map is defined as follows:
\begin{small}

\begin{eqnarray*}
 x_{n+1}=
\left\lbrace \begin{array}{ll}
LT_1(r,x_n):=rx_n(1-x_n)\\
~~~~~~~~~~~~~~~~~~~~+\frac{4-r}{2}x_n,&~~~~~~~x_n<0.5, \\
LT_2(r,x_n):=rx_n(1-x_n)\\
~~~~~~~~~~~~~~~~~~~~+\frac{4-r}{2}(1-x_n),&~~~~~~~x_n\geq 0.5.
\end{array}\right.
\end{eqnarray*}

\end{small}
By a combination and transformation of these maps,
a hybrid chaos map is introduced as follows\cite{6}:
\begin{eqnarray*}
&& x_{n+1}:=
 HCM1(r,x_n)=\\
&&\left\lbrace \begin{array}{ll}
r\Big( \sin \big(LT_1(r,x_n)\big)+\cot(x^2_n)\\
~~~~~~~~~~~~+S\big(r,LT_1(r,x_n)\big)\Big),~~~~~~~~~~x_n<\frac{1}{3}, \\
(r+1)\Big(\sin\big(LT_2(r,x_n)\big)+\cot(x_n)\\
~~~~~~~~~~~~+10^5\sqrt{x_n}\Big),~~~~~~~~~~~~~~~~~~~~~~~x_n\geq\frac{1}{3}.
\end{array}\right.
\end{eqnarray*}
Also, in \cite{5}, 2D Hybrid chaos map is expressed as follows:

\begin{eqnarray}
&&x _{n+1}:=HCM2x(r,x_n,y_n)=\nonumber\\
&&\left\{%
\begin{array}{ll}
\omega^x_1f^x_1 \circ F^x_1(r,x_n)+\alpha^x_1 g^x_1(r,x_n,y_n)\\
~~~~~~~~~~~~+h_1^x\big(\frac{(\beta^x_1-r)x_n}{2}\big)~mod~1,~~~~~~~~~y_n<0.5,\\
\\
\omega^y_2f^y_2 \circ F^x_2(r,x_n)+\alpha^x_2 g^x_2(r,x_n,y_n)\\
~~~~~~~~~~~~+h_2^x\big(\frac{(\beta^x_2-r)(1-x_n)}{2}\big)~mod~1,~~~y_n\geq0.5,\\
\end{array}%
\right.
\end{eqnarray}

\begin{eqnarray}
&&y_{n+1}:=HCM2y(r,x_n,y_n)=\nonumber\\
&&\left\{%
\begin{array}{ll}
\omega^y_1f^y_1 \circ F^y_1(r,y_n)+\alpha^y_1g^y_1(r,\zeta,y_n)\\
~~~~~~~~~~~~+h_1^y\big(\frac{(\beta^y_1-r)\zeta}{2}\big)~mod~1,~~~~~~~~~~~\zeta<0.5,\\
\\
\omega^y_2f^y_2 \circ F^y_2(r,y_n)+\alpha^y_1g^y_2(r,\zeta,y_n)\\
~~~~~~~~~~~~+h_2^y\big(\frac{(\beta^y_2-r)(1-\zeta)}{2}\big)~mod~1,~~~~~\zeta\geq0.5.\\
\end{array}%
\right.
\end{eqnarray}
where
$$\zeta=x_n ~~ or ~~ x_{n+1},$$
$(\omega^k_l,\alpha^k_l , \beta^k_l)$, $(f^k_l, h^k_l)$ and $g^k_l$ for $(k=x,y ~\& ~l=1,2)$ have been respectivly considered as weights, combination maps and transfer maps. Also $F^k_l$ have been arbitrary chosen Logistic or Sine maps.

\subsection{The new chaotic system}\label{sec1}
By takeing advantages of the above hybrid chaos maps, we aim to develop a new type of chaotic map.
The
process of creating proposed hybrid system is shown in Fig. \ref{flo}. To
create the proposed hybrid chaos system, as we can see from the figure,
two hybrid chaos maps are used to product the inputs of third composite box. Within this box, numerical
values are entered and by using the combination and transfer operations,
the output value is generated.
This map is structured as follows:
\line(1,0){230}
$$\rho_n =HCM1(r_1,x_n),$$
$$\rho_{n+1} =HCM1(r_1,\rho_n),$$
$$\tau^1_n= HCM2x(r_2,\rho_n,\rho_{n+1}),$$
$$\tau^2_n= HCM2y(r_2,\rho_n,\rho_{n+1}),$$
$$ x_{n+1}:= \gamma \varphi_1 o\varphi_2(\tau^1_n,\tau^2_n),$$
\line(1,0){230}\\
\noindent where $\gamma$ is a constant and $\varphi_1$ and $\varphi_2$ are appropriate continuous functions.

\begin{figure}
\centering
\includegraphics*[height=4cm, width=8cm]{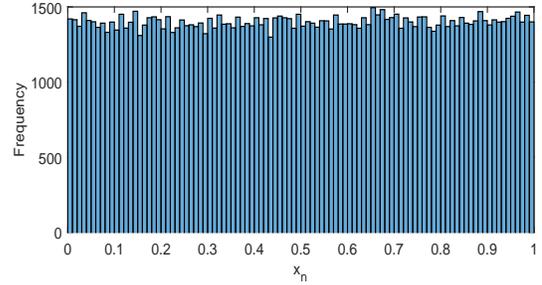}
\caption{Histogram plot of the proposed system by $x_0=0.03$, $r_1=0.01$ and $r_2=0.3$.
}\label{H1}
\end{figure}
\begin{figure*}[!t]
\centering
\subfloat[$r_1=0.01,r_2=0.3$]{\includegraphics[width=3.2in]{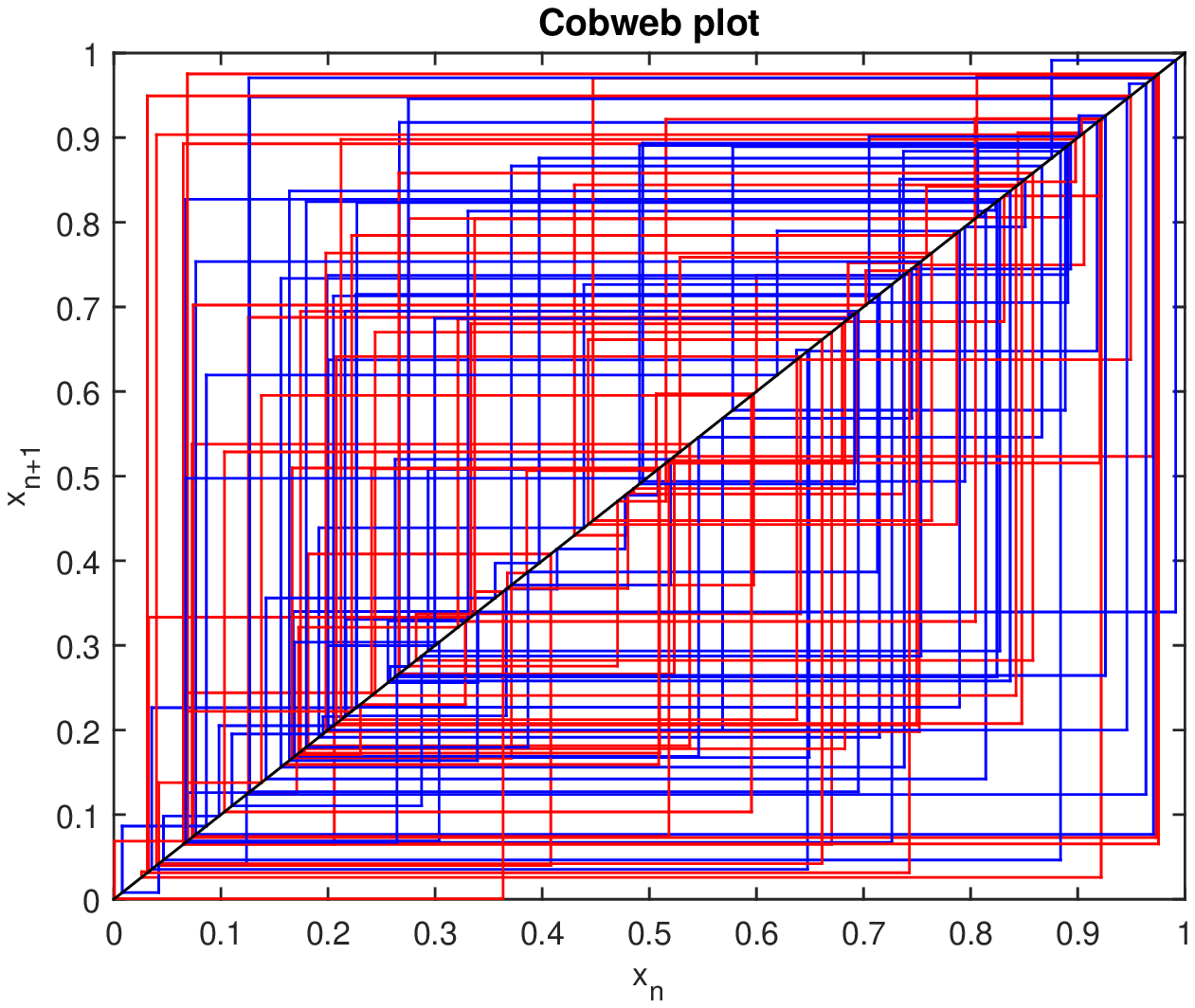}%
}
\hfil
\subfloat[$r_1=0.3,r_2=0.01$]{\includegraphics[width=3.2in]{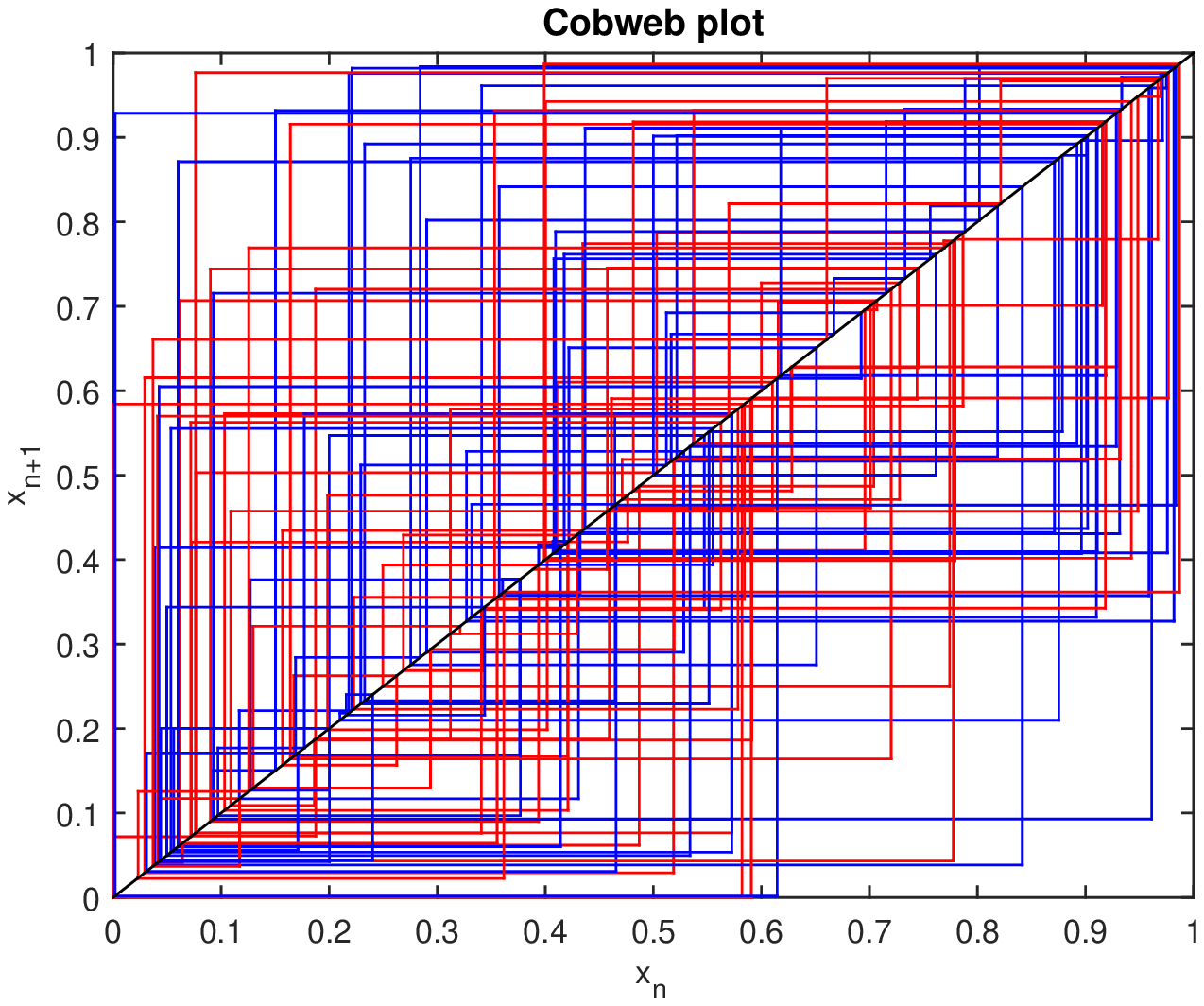}}%
\caption{Cobweb plots for different values of $r_1$ and $r_2$ with $x_0=0.2$ for blue color and $x_0=0.6$ for red color and.}
\label{co1}
\end{figure*}

The histogram of the proposed system for 140,000 points is shown in Fig. \ref{H1}.
By using this figure, it is can been seen that the histogram is almost flat. Another suitable diagram to
investigate the behavior of a chaos system is the cobweb plot. Cobweb plots
for different values of $r_1$ and $r_2$ are given in Fig. \ref{co1}. The obtained results show that the output states
does not converge to a special point.

\section{Performance analysis of proposed chaotic map}
The proposed chaotic map exhibit complex chaotic behaviors. To demonstrate this property, we estimate the chaos performance of chaotic map produced above. The tests are performed in terms of initial state sensitivity, bifurcation diagrams and Lyapunov exponents.


\subsection{Initial state sensitivity}
\begin{figure}
\centering
\includegraphics*[height=4cm, width=8cm]{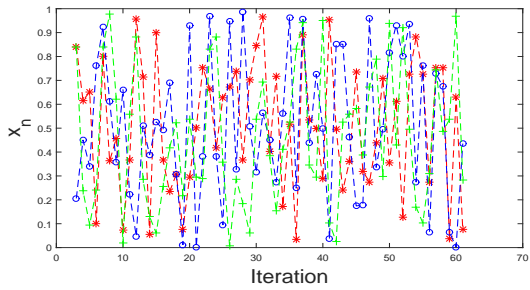}
\caption{Three outputs of the proposed system: (red) $x_0=0.6, r_1=0.1, r_2=0.3$ (blue) $x_0=0.6+10^{-16}, r_1=0.1, r_2=0.3$. (green) $x_0=0.6, r_1=0.1+10^{-16}, r_2=0.3$
}\label{it1}
\end{figure}

To check the system sensitivity to initial points, very small changes as $+10^{-16}$ have been
made to the input data. The output values for different initial points are shown in Fig. \ref{it1}.
It can be seen that the output values are changed by a small
change in the initial value.

\begin{figure*}[!t]
\centering
\subfloat[$x_0=0.5$]{\includegraphics[width=3.2in]{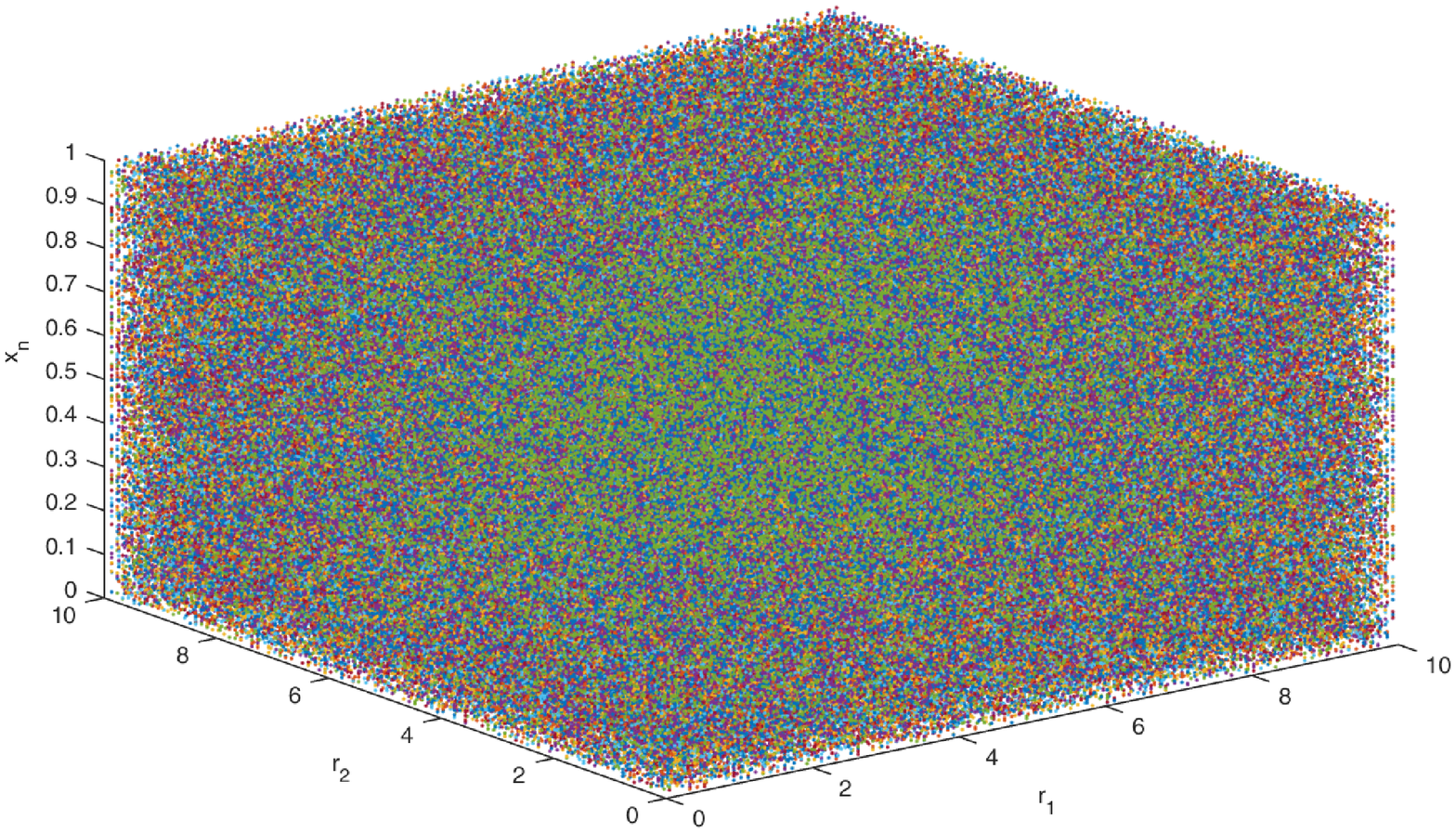}%
}
\hfil
\subfloat[$x_0=0.2$]{\includegraphics[width=3.2in]{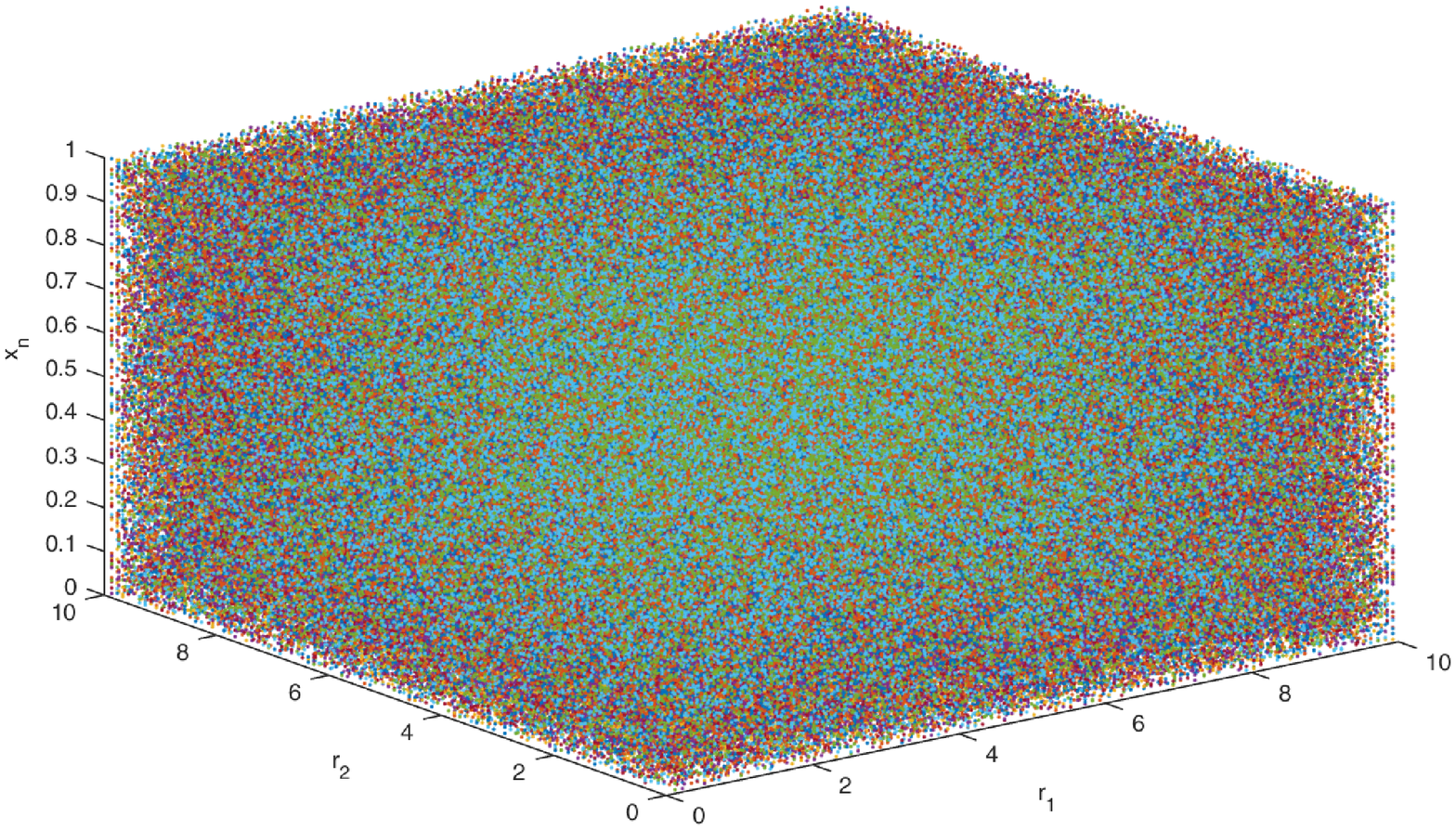}}%
\caption{Bifurcation diagrams of the proposed system for different values of $x_0$.}
\label{bi1}
\end{figure*}

\subsection{Bifurcation diagram}
The bifurcation diagram of a dynamical system provides a visualized method to exhibit the chaotic system behaviors. Fig. \ref{bi1} presents the bifurcation diagrams of the proposed map for two initial values $x_0=0.2$ and $x_0=0.5$. The existing maps, namely the Logistic, Sine, and Tent maps, have fixed or periodic points in most parameter settings. Moreover, their output states are only distributed in a small area. However, the generated chaotic map exhibit complicated behaviors in all parameter ranges, and its output states are randomly distributed in the entire plane, indicating that it has robust chaotic behaviors and its outputs are more random.
\subsection{Lyapunov exponent analysis}
Another tool for studying chaos system is Lyapunov exponent.
The relationship between Lyapunov exponent and chaotic system
can be considered in the following theorem.

\textbf{Theorem}
If at least one of the average Lyapunov exponents is positive, then the system is chaotic; if the average
Lyapunov exponent is negative, then the orbit is periodic and when the average Lyapunov exponent is zero,
a bifurcation occurs\cite{10}.

\begin{figure}
\centering
\includegraphics*[height=4cm, width=8cm]{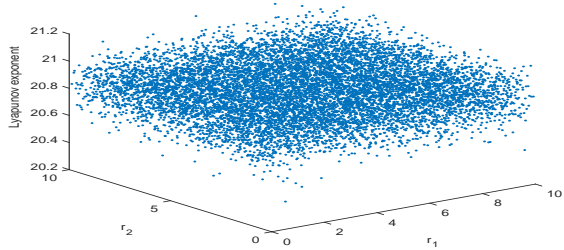}
\caption{Lyapunov exponent results for proposed
hybrid chaotic system with $x_0=0.5$.
}\label{ly1}
\end{figure}

Thus, a dynamical system is considered to exhibit chaotic behaviors if it can obtain a positive Lyapunov exponent. A larger Lyapunov exponent means that the trajectories diverge faster, indicating superior chaos performance. Fig. \ref{ly1}
shows that the numerical values for lyapunov exponent are
positive then this test indicates that neighboring trajectories diverge.

\begin{figure}[!t]
\centering
\subfloat{\includegraphics[width=3in]{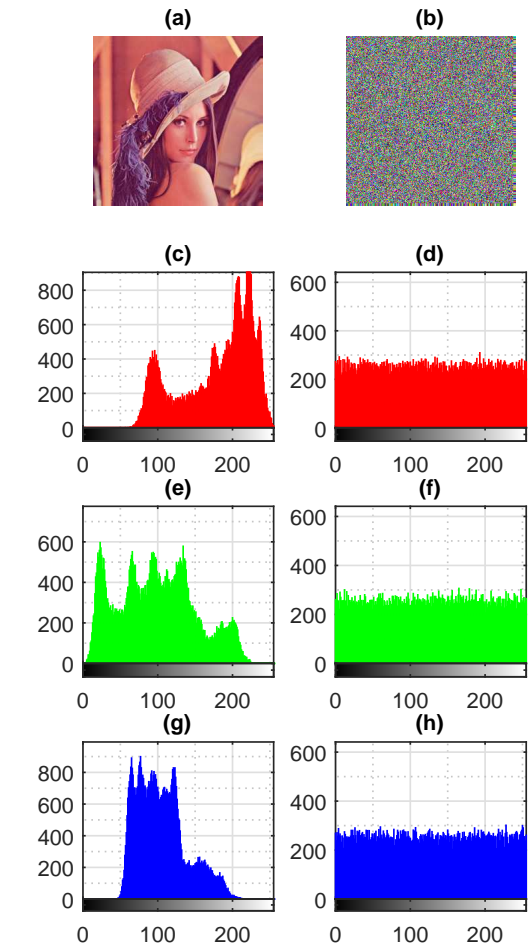}%
}
\vspace*{-1cm}
\caption{Original lena image, encryption results and histograms: (a) lena
original image; (b) encrypted lena image; (c),(e) and (g) histograms of original lena; (d), (f) and (h) histograms of encrypted lena.}
\label{l1}
\end{figure}

\section{ Encryption results}
To prove resistance and sensitivity of the new system in security algorithms, we apply
outputs of chaos sequence in the image encryption algorithm suggested in \cite{7}. The proposed map is replaced on the encryption algorithm.
The obtained results for the encryption algorithm utilizing the generated chaotic map are given in Fig. \ref{l1}.

The resistance of an encrypted image against differential attacks can be measured by the Number of
Pixels Change Rate (NPCR) and the Unified Average Changing Intensity (UACI). When these values are close to
$100\%$ and $33.33\%$ respectively, then encryption algorithm has high sensitivity of changing of plain image.
In order to test the proposed algorithm, the results obtained for NPCR and UACI have been compared with the
algorithm \cite{7} in Table \ref{t1}.
\begin{table}[!t]
\caption{Comparison of NPCR and UACI values for Lena $(256\times 256 \times 3)$ using the proposed and \cite{7} maps.}\label{t1}\begin{center}\begin{tabular}{ccccc}
 \hline \\ $Measure$ &&&$Encrypted~lena$\\
  \cline{3-5}
 \\
& &&Ref.\cite{7}&$Proposed~map$\\
\hline
\\
&&R& 99.5804& 99.6368\\
\\
NPCR&&G&99.6628&99.6277\\
\\
&&B&99.5865&99.6399\\
\\
Average&&&99.6099&99.6348\\
\hline
\\
&&R&33.4456&33.3907\\
\\
UACI&&G& 33.6561& 33.4945\\
\\
&&B&33.5512&33.5559\\
\\
Average&&&33.5509&33.4524\\
\hline \end{tabular}\end{center}\end{table}
\section{Conclusion}
This paper has proposed a new chaotic map. It was derived from the hybrid chaos and 2D hybrid chaos systems utilizing the Sine and Logistic maps. Several assessment methods, including the Lyapunov exponents, initial states sensitivity, bifurcation diagram and etc, have been used to evaluate the chaotic performance of the chaos map. Analysis and evaluation results have shown that this map has the wider chaotic range, better ergodicity and hyperchaotic property, and that it has better chaotic performance than existing chaotic maps.


%

\ifCLASSOPTIONcompsoc


\ifCLASSOPTIONcaptionsoff
  \newpage
\fi

\end{document}